\documentclass[usenatbib,useAMS,usedcolumn]{mn2e}

\usepackage{epsfig}
\usepackage{times}
\pdfoutput=1

\newcommand{\arcm}{\hbox{$^\prime$}}

\newcommand{\chandra}{\emph{Chandra}}
\newcommand{\xmm}{\emph{XMM-Newton}}
\newcommand{\xmms}{\emph{XMM}}
\newcommand{\asca}{\emph{ASCA}}

\newcommand{\arcs}{\mbox{\arcm\arcm}}

\newcommand{\Zsol}{\ensuremath{Z_{\odot}}}

\newcommand{\Msol}{\ensuremath{M_{\odot}}}

\newcommand{\s}{\ensuremath{\mbox{~s}}}
\newcommand{\ps}{\ensuremath{\s^{-1}}}
\newcommand{\cm}{\ensuremath{\mbox{~cm}}}
\newcommand{\pcmsq}{\ensuremath{\cm^{-2}}}
\newcommand{\km}{\ensuremath{\mbox{~km}}}

\newcommand{\kmps}{\ensuremath{\km \ps}}
\newcommand{\mJy}{\ensuremath{\mbox{~mJy}}}
\newcommand{\bm}{\ensuremath{\mbox{~b}}}
\newcommand{\pb}{\ensuremath{\bm^{-1}}}
\newcommand{\mJypb}{\ensuremath{\mJy \pb}}

\newcommand{\Ovii}{\mbox{{O{\small VII}}}}

\newcommand{\Ho}{\ensuremath{H_\mathrm{0}}}

\newcommand{\Dtf}{\ensuremath{D_{\mathrm{25}}}}

\newcommand{\gtsim}{\,\rlap{\raise 0.5ex\hbox{$>$}}{\lower 1.0ex\hbox{$\sim$}}\,}

\begin{document}

\title[ 
jet-driven enrichment in AWM~4
] 
{ 
A deep \textit{Chandra} observation of the poor cluster AWM~4 -- II. The role of the radio jets in enriching the intra--cluster medium.
}

\author[E. O'Sullivan et al.]  {Ewan O'Sullivan\footnotemark[1]$^{1,2}$,
  Simona Giacintucci$^{2,3}$, Laurence P. David$^{2}$, Jan
  M. Vrtilek$^{2}$ \newauthor and Somak Raychaudhury$^{1}$\\
  $^{1}$ School of Physics and Astronomy, University of Birmingham,
  Edgbaston, B15 2TT, UK \\
  $^{2}$ Harvard-Smithsonian Center for Astrophysics, 60 Garden Street, Cambridge, MA 02138 \\
  $^{3}$ INAF -- Instituto di Radioastronomia, via Gobetti 101, 40129
  Bologna, Italy }

\date{Accepted 2010 October 2.  Received 2010 September 30; in original form 2010 June 24}

\pagerange{\pageref{firstpage}--\pageref{lastpage}} \pubyear{2010}

\maketitle

\label{firstpage}

\begin{abstract} 
  We use a \chandra\ observation of the poor cluster AWM~4 to map the
  temperature and abundance of the intra--cluster medium, so as to examine
  the influence of the central radio galaxy on its environment.  While the
  cluster core is generally enriched to near--solar abundances, we find
  evidence of super--solar abundances correlated with the radio jets,
  extending $\sim$35~kpc from the core of the central dominant galaxy
  NGC~6051 along its minor axis. We conclude that the enriched gas has been
  transported out of the central galaxy through the action of the radio
  source.  We estimate the excess mass of iron in the entrained gas to be
  $\sim$1.4$\times10^6$\Msol, and find that this can be produced in the
  core of NGC~6051 within the timescale of the AGN outburst. The energy
  required to transport this gas to its current location is
  $\sim4.5\times10^{57}$~erg, a significant fraction of the estimated total
  mechanical energy output of the AGN, though this estimate is dependent on
  the degree of enrichment of the uplifted gas.  The larger near--solar
  abundance region is also compatible with enrichment by metals mixed
  outward from NGC~6051 over a much longer timescale.

\end{abstract}


\footnotetext[1]{E-mail: ejos@star.sr.bham.ac.uk}

\section{Introduction}
\label{sec:intro}

Heating by active galactic nuclei (AGN) is currently regarded as the most
likely mechanism preventing excessive cooling of the hot intra--cluster
medium (ICM) in galaxy groups and clusters \citep[e.g.,][and references
therein]{McNamaraNulsen07,PetersonFabian06}. Cool core clusters typically
have a central giant elliptical or cD galaxy, 70-100 per cent. of which
host radio sources \citep{Burns90,Mittaletal09}. Accretion of cooling gas
onto the central supermassive black hole of the central galaxy is believed
to fuel outbursts which then reheat the cooling material.

Cool core clusters are also known to have strongly centrally peaked
abundance profiles \citep[e.g.,][]{Bohringeretal04,DeGrandiMolendi04}, as
are many X-ray bright galaxy groups \citep[e.g.,][]{RasmussenPonman07}. The
location of the dominant elliptical at the centre of this peak suggests
that it is probably the source of these metals, and observed metal masses
are found to be consistent with the quantities expected from supernovae and
stellar winds. In particular, the iron mass profiles of systems across a
wide range of mass are consistent with long-term enrichment primarily by
SNIa in the central dominant galaxy
\citep{Bohringeretal04,RasmussenPonman07,Werneretal08}.

The iron mass profiles of cool core systems are found to be more extended
than the light distribution of the central galaxy, indicating that metals
are transported outward from the cluster core
\citep[e.g.,][]{DavidNulsen08,RasmussenPonman09}. The central AGN is
clearly a strong candidate to drive this process, and numerous studies have
used metallicity measurements to constrain models of AGN--driven gas
motions
\citep[e.g.,][]{Rebuscoetal05,Rebuscoetal06,DavidNulsen08,Xiangetal09}.
Evidence of enrichment associated with the radio jets and lobes of central
radio galaxies has been found in nearby X--ray bright clusters
\citep[e.g.,][]{Sandersetal04,Simionescuetal08,Simionescuetal09,Kirkpatricketal09},
suggesting that AGN--driven outflows or entrainment is a viable mechanism
for transporting enriched gas to large radii. However, with unambiguous
examples identified only in a handful of clusters, it is as yet unclear how
common this process is, and the identification and characterisation of
outflows in other systems is a necessity.

In this paper we construct temperature and abundance maps of the poor
cluster AWM~4, so as to investigate the effect of an ongoing AGN outburst
on the structure of the intracluster medium (ICM) and in particular the
metallicity distribution. The central dominant elliptical galaxy, NGC~6051,
hosts a powerful radio source, 4C+24.36, which has been characterised in
detail using deep multi-frequency observations from the Giant Metrewave
Radio Telescope \citep{Giacintuccietal08}. We have previously examined the
temperature and abundance structure of the cluster using \xmm\
\citep[hereafter referred to as OS05]{OSullivanetal05_special}, and the
large-scale temperature and abundance profiles of AWM~4 have also been
determined from these data \citep{Gastaldelloetal08}. More recently we have
used the \chandra\ data described in this paper to place limits on the
properties of the radio galaxy, and in particular to determine the
mechanical power required to form the jets and lobes, the nature of the
plasma in the lobes, and the properties of the central galactic corona
\citep[hereafter referred to as paper I]{OSullivanetal10a_special}. 

Both the ICM and galaxy distribution of the cluster appear relaxed, with a
concentration of early-type galaxies toward the cluster core
\citep{KoranyiGeller02}. NGC~6051 shows no indication of having undergone
recent interactions \citep{Schombert87} and is the most luminous galaxy in
the cluster by a significant factor, with a magnitude difference above the
second ranked galaxy of $M_{12}$=1.6 (SDSS $g$ band). Giacintucci et al.
show 4C+24.36 to be a wide--angle--tail radio galaxy with inner jets
oriented close to the plane of the sky, probably moving southward with a
velocity $\la$120\kmps.  Throughout the paper we assume \Ho=70,
$\Omega_M=0.3$, and $\Omega_{\Lambda}=0.7$. The redshift of the cluster is
taken to be 0.0318, and we assume angular size and luminosity distances of
$D_A$=130.9~Mpc and $D_L$=139.3~Mpc respectively.

\section{Observations and Data Reduction}
\label{sec:obs}

AWM~4 was observed by the \chandra\ ACIS instrument during Cycle~9 on 2008
May 18-19 (ObsId 9423), for $\sim$80~ks. A summary of the \chandra\ mission
and instrumentation can be found in \citet{Weisskopfetal02}. The S3 CCD was
placed at the focus of the telescope and the instrument operated in very
faint mode. We reduced the data from the pointing using CIAO 4.1.2 and
CALDB 4.1.2 following techniques similar to those described in
\citet{OSullivanetal07} and the \chandra\ analysis
threads\footnote{http://asc.harvard.edu/ciao/threads/index.html}.  The
level 1 events files were reprocessed, very faint mode filtering applied,
bad pixels and events with \asca\ grades 1, 5 and 7 were removed, and the
cosmic ray afterglow correction was applied. The data were corrected to the
appropriate gain map, the standard time-dependent gain and charge-transfer
inefficiency (CTI) corrections were made, and a background light curve was
produced.  The final cleaned exposure time was 74.5~ks.

Identification of point sources on S3 was performed using the \textsc{ciao}
task \texttt{wavdetect}.  Source ellipses were generated with axes of
length 4 times the standard deviation of each source distribution. These
were then used to exclude sources from all further analysis. A source was
detected coincident with the peak of the diffuse X-ray emission; this was
not excluded.

Spectra were extracted and responses created for each region following the
methods used by the \textsc{ciao} \texttt{specextract} command. Background
spectra were drawn from the standard period D set of CTI-corrected ACIS
blank sky background events files in the \chandra\ CALDB, normalised to
match the 9.5-12.0 keV count rate in the target observation. Very faint
mode screening was applied to the background data. A slight excess of soft
emission, compared to the source data, is observed in the background
datasets, mainly below 0.5 keV. This is not unexpected, as
  the soft X--ray background arises largely from hot gas in the Milky Way
  and from coronal emission associated with solar wind interactions, and is
  both spatially and temporally variable \citep[e.g.,][]{KuntzSnowden00,
    Snowdenetal04}. There are also indications that the spectral shape of
  the background of the ACIS-S3 CCD has changed since the creation of the
  period D background files (c.f. the ACIS background
  cookbook\footnote{http://asc.harvard.edu/contrib/maxim/acisbg/COOKBOOK}),
  which could contribute to the disagreement at low energies. As discussed
  in paper I, experimentation with fitting different energy bands suggests
  that ignoring energies below 0.7~keV produces results consistent with
  those found using \xmm\ in OS05, and these results are relatively
  insensitive to changes in the absorbing hydrogen column. We therefore
  ignore counts outside the 0.7-7~keV band in our fits.

 We also examined the possibility of modelling the soft background
  excess, following the technique described by \citet{Vikhlininetal05}, in
  which the background subtracted spectrum of a source--free region is
  fitted using a zero redshift unabsorbed plasma model with temperature
  $\sim$0.2~keV, whose normalisation is allowed to be negative. Since AWM~4
  entirely fills the S3 CCD, we extracted spectra from the S1 CCD, which is
  also back illuminated and has a similar response. We ignored energies
  below 0.5~keV, where instrument calibration is less reliable. We found
  that a 0.2~keV apec model alone, or in combination with a slightly harder
  0.3-0.5~keV apec model is insufficient to model the residual, owing to a
  line feature at $\sim$0.55~keV. This probably corresponds to \Ovii\
  emission, and we therefore added a Gaussian with negative normalisation
  to model this component. We then tested the effects of this model,
  appropriately scaled, on fits to spectra from a number of regions on the
  S3 CCD. The correction was moderately successful, but did not completely
  remove residuals, suggesting that our model may be too simple or that
  variations in the background datasets for S1 and S3 are significant. We
  note that we did not use the combined S1/S3 background file, since this
  has poorer statistics than the S3--only dataset. The effect of the
  correction on fit parameters was to marginally raise the temperature and
  in some regions the abundance, and to increase the uncertainty on all
  parameters. However, the abundances agreed with those measured in the
  0.7-7~keV within the uncertainties, and there was no evidence that the
  abundance distribution would be altered by using the correction. 

Since this analysis was completed, newer blank-sky background datasets
(period E), more appropriate for use with this observation, have been
released.  We have tested the effect of using the period E backgrounds on
our results by reprocessing the data, re-extracting the map spectra with
the newer background, and refitting them. While there are differences in
map pixel values, these are within the uncertainties, and we conclude that
our original background treatment is sufficient for our goals.  Spectral
fitting was performed in XSPEC 11.3.2ag. Abundances were measured relative
to the abundance ratios of \citet{GrevesseSauval98}.  We assumed a
  galactic hydrogen column of 0.05$\times10^{22}$\pcmsq\ \citep[from the
  \textsc{ftools} task \texttt{nh}, based on the data
  of][]{Kalberlaetal05}, and 90\% uncertainties on fitted parameters were
  estimated.  Spectra were grouped to 20 counts per bin.

\section{Spectral Maps}
\label{sec:maps}

\begin{figure*}
\centerline{
\includegraphics[width=0.333\textwidth, viewport=36 41 576 752]{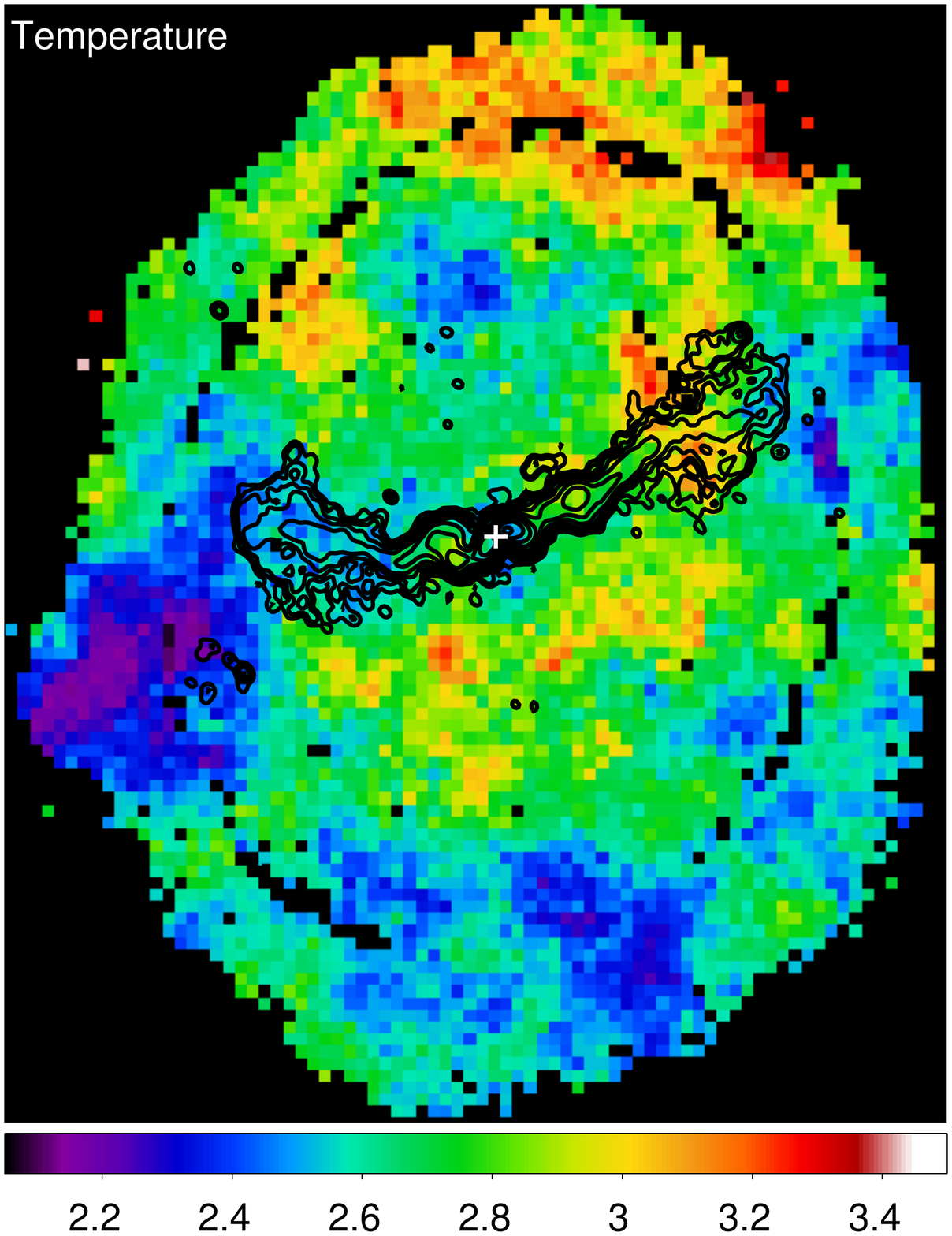}
\includegraphics[width=0.333\textwidth, viewport=36 41 576 752]{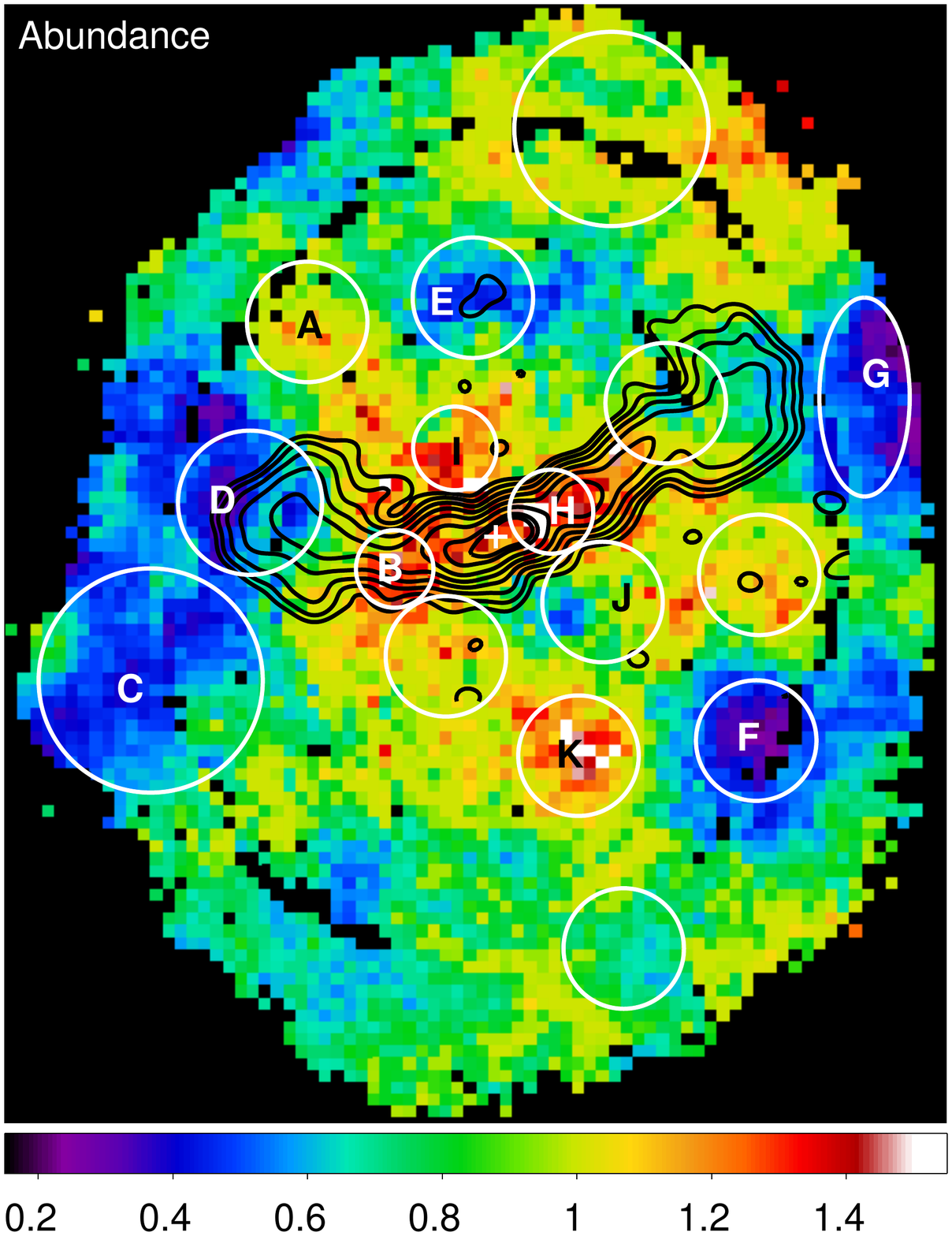}
\includegraphics[width=0.318\textwidth,viewport=36 -10 577 721]{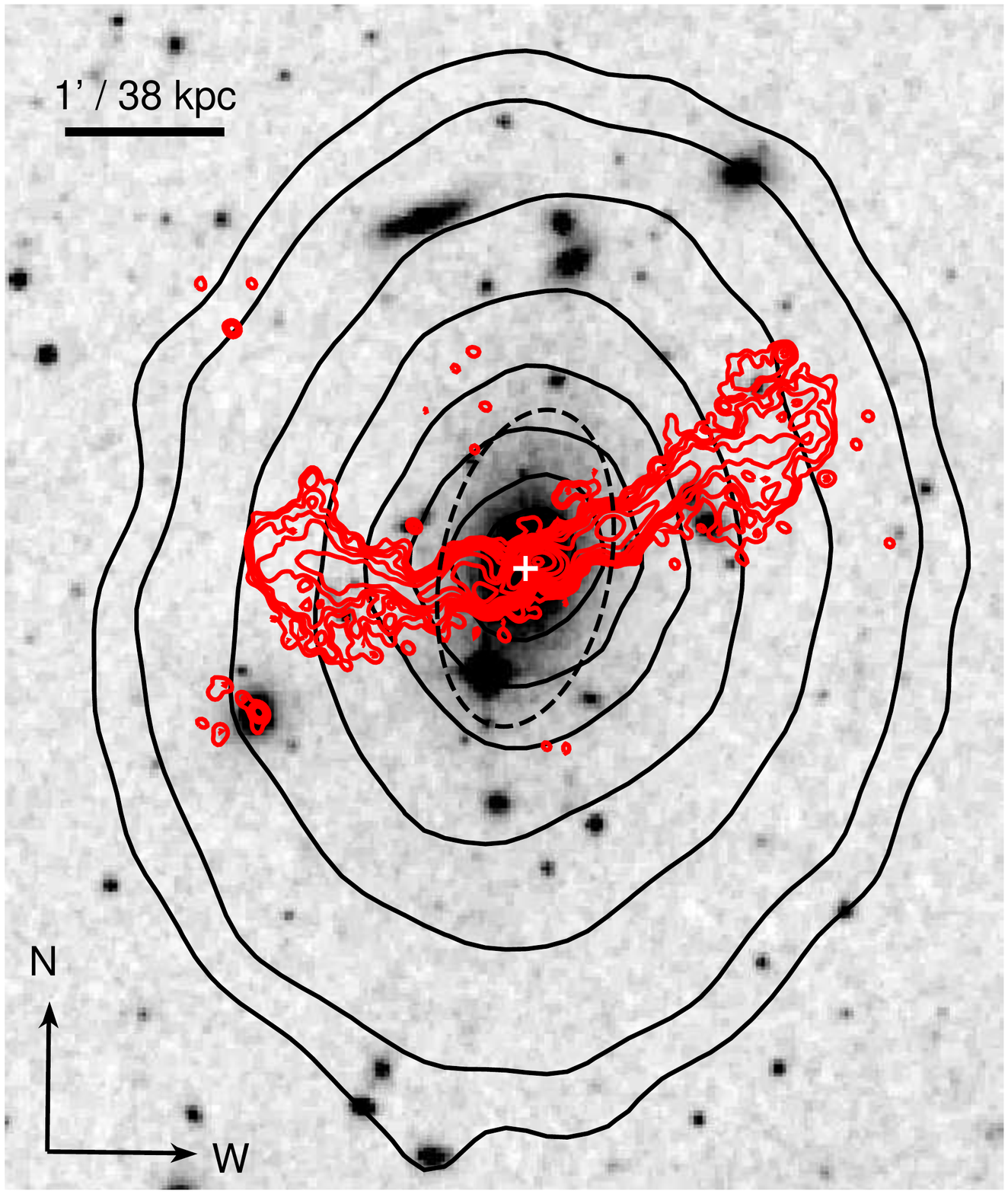}
}
\caption{\label{fig:maps} Temperature (\textit{left}) and abundance
  (\textit{centre}) maps of AWM~4 in units of keV and \Zsol\ respectively.
  Black pixels indicate regions with insufficient surface brightness,
  excluded point sources, or temperature uncertainties exceeding 15\%. GMRT
  610~MHz contours (starting at 0.2\mJypb\ and spaced by factor 2) are
  overlaid on the temperature map, 235~MHz contours (starting at 2.4\mJypb\
  and spaced by factor 2) on the Abundance map. White crosses indicate
  the position of the radio core. Circular and elliptical regions on the
    abundance map were used to compare the maps consistency with normal
    spectral fits. Lettered regions are discussed in the text. A Digitized
  Sky Survey $B_j$-band image (\textit{right}) is shown for comparison,
  with the same scale and alignment as the other images.  The approximate
  \Dtf\ contour of NGC~6051 is marked by a dashed ellipse, 610~MHz radio
  contours are overlaid in red and smoothed X--ray contours (starting at
  0.075 ct. arcsec$^{-1}$, 0.3-2~keV, and spaced by factor $\sqrt{2}$) are
  overlaid in black.  }
\end{figure*}

We follow a process similar to that described in OS05 to create adaptively
binned 2-D maps of temperature and abundance from the \chandra\
observation.  Our mapping technique uses a fixed map pixel size, but allows
the spectral extraction region associated with each pixel to vary
adaptively. These regions are typically larger than the map pixels, so the
resulting pixel values are not independent. The maps are thus analogous to
adaptively smoothed images, in which the pixel scale is fixed, but the
effective resolution varies over the image. For this reason, we emphasise
that we use the maps only as a diagnostic tool, to identify interesting
features and regions for further investigation.

\begin{figure}
\centerline{\includegraphics[width=\columnwidth,viewport=35 193 577 600]{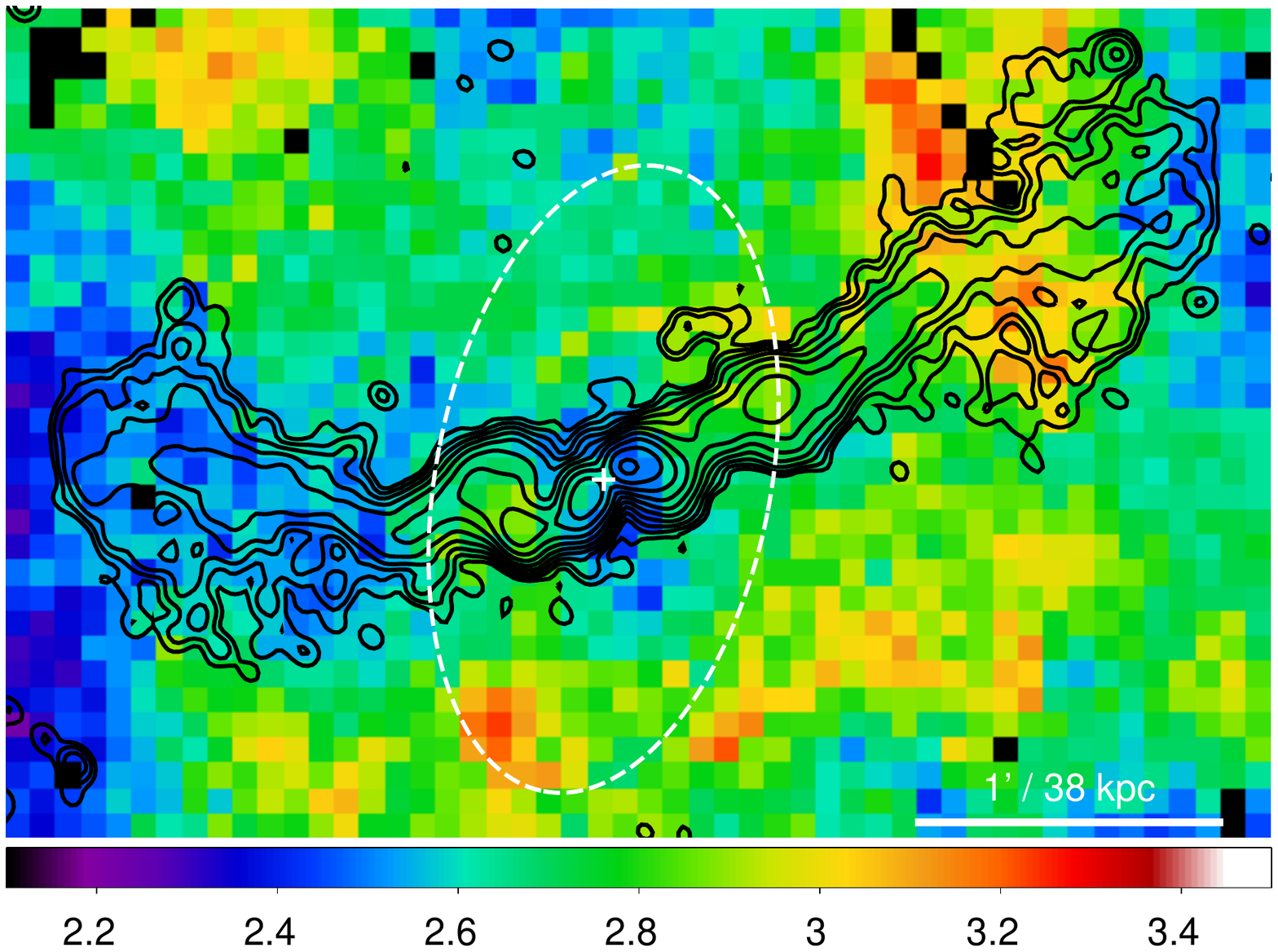}}
\caption{\label{fig:Tzoom} Enlarged view of the temperature map in the
  region surrounding the radio source. GMRT 610~MHz contours (defined as
  described in Fig.~\protect\ref{fig:maps}) are marked in black, and
  the approximate \Dtf contour is marked by a dashed white ellipse. The white cross marks the position of the radio core.}
\end{figure}

We define the map pixels to be square with side length 4.93\arcs\ (10 ACIS
physical pixels). We require the associated spectra to contain 1600 net
counts in the range 0.7-7.0~keV, extracted using circular regions with
radii in the range $\sim$14-53\arcs. Spectra were fitted with an absorbed
APEC thermal plasma model. Any pixel with an uncertainty in temperature of
$>$15 per cent was excluded as unreliable. The final maps are shown in
Figure~\ref{fig:maps}, and an enlarged view of the central part of the
temperature map is shown in Figure~\ref{fig:Tzoom}.

Several interesting features are visible in the spectral maps. The central
small-scale cool core or galactic corona is visible in the temperature map
as a cool region coincident with the centre of NGC~6051 and the origin of
the radio jets.  There is some indication of cool temperatures along the
eastern jet and particularly in the eastern lobe. A weak cavity in the ICM
is detected coincident with the eastern lobe (see paper I) and the cool
temperatures may arise because of the reduced quantity of high--temperature
plasma along the line of sight. The azimuthally averaged temperature
profile of AWM~4 declines at radii beyond $\sim$200~kpc ($\sim$5.25\arcm),
falling to 1.5-2~keV at 400~kpc \citep{Gastaldelloetal08}, so a cavity in
the hot gas will result in a higher fraction of emission from this cool
material in the spectra extracted for this region.  In contrast, a hotter
region is located where the eastern jet bends sharply north and then
returns to its eastward course. The western jet passes through higher
temperature regions, particularly where the jet enters the lobe, and the
highest temperature material is to the north of the jet at the base of the
lobe.  High temperatures are also seen in the radio--bright knot in the
western jet.  Increased temperatures could indicate heating by shocks
driven by the jet.  Features such as the western knot could indicate
obstruction of the jet by ICM plasma, while the correlation with bends in
the eastern jet suggests either obstruction or outward motion of the jet
``wiggles''.  However, the high temperature region in the western lobe is
poorly correlated with the radio emission and is likely to be unrelated.

More generally, in the inner part of the cluster temperatures are higher
south of the jet than to the north. At larger radii this trend is reversed,
with high temperatures in the north and cooler material to the south, east
and west. There is no correspondence between the stellar envelope of
NGC~6051 and the temperature structure.

The most notable features in the abundance map (Figure~\ref{fig:maps},
middle panel) are a general roughly solar abundance region in the centre of
the map, which is asymmetric and clumpy, indicating uneven enrichment of
the ICM. Regions of super-solar abundance extend along the jets to the
radius of the western knot and the base of the eastern lobe, with some
extension to the north. The solar abundance region is somewhat more
extended south of the jets, but is less consistent, with patches of both
high and low abundance. At larger radii the abundances decline, but there
is considerable variation, from $\sim$0.4\Zsol\ regions in the east and
west, to solar abundances in the north and south.

The correlation between the super--solar abundance region and the radio
jets suggests that enriched material is being entrained outward from the
core of NGC~6051. A branch or clump of high abundances also extends north
or northeast from the central galaxy. This may indicate a trail of material
left behind the galaxy, since the bending of the radio jets suggests that
NGC~6051 is moving south.  Alternatively it could indicate that the
entrainment of gas to the east is less closely confined around the jet than
is the case on the west.  Neither the highest abundance features nor the
larger near-solar region correlate with the stellar structure of NGC~6051.
The high abundance feature extends roughly across the minor axis of the
galaxy, but is more extended that the \Dtf\ ellipse.

Comparison of the maps with galaxies in the field of view shows no clear
correlations. IC~4588, an early--type galaxy at redshift 0.051 falls at the
western edge of the large cool, low abundance region to the southeast of
the eastern radio lobe (region 1 in Figure~\ref{fig:maps}).  It is possible
that the cool material is associated with the galaxy, perhaps as part of a
galaxy group.  \citet{KoranyiGeller02} find a small number of galaxies at
approximately the same recession velocity. However, there is no clear
surface brightness structure in the region, and there are insufficient
counts to allow us to identify any additional spectral components. An
apparent radio source coincident with IC~4588 is seen in the 610~MHz
contours, but comparison with the available GMRT and VLA maps at other
frequencies suggests that while there is a source at this position, its
apparent extension is the result of a noise feature.

\subsection{Metal enrichment along the jets}
\label{sec:metals}
Figure~\ref{fig:Zerr} shows the map of best-fitting abundance values and
the associated 90 per cent upper and lower bound maps. The central
abundance feature which correlates with the jets is clear in all three
maps. Maps of the fit statistic show variation across the field, but do not
appear correlated with the temperature or abundance maps. This suggests
that the apparent features are not the product of poor spectral fits in
particular regions. We test this conclusion more throughly in
Section~\ref{sec:bias}.

\begin{figure}
\centerline{\includegraphics[width=\columnwidth,viewport=35 177 577 616]{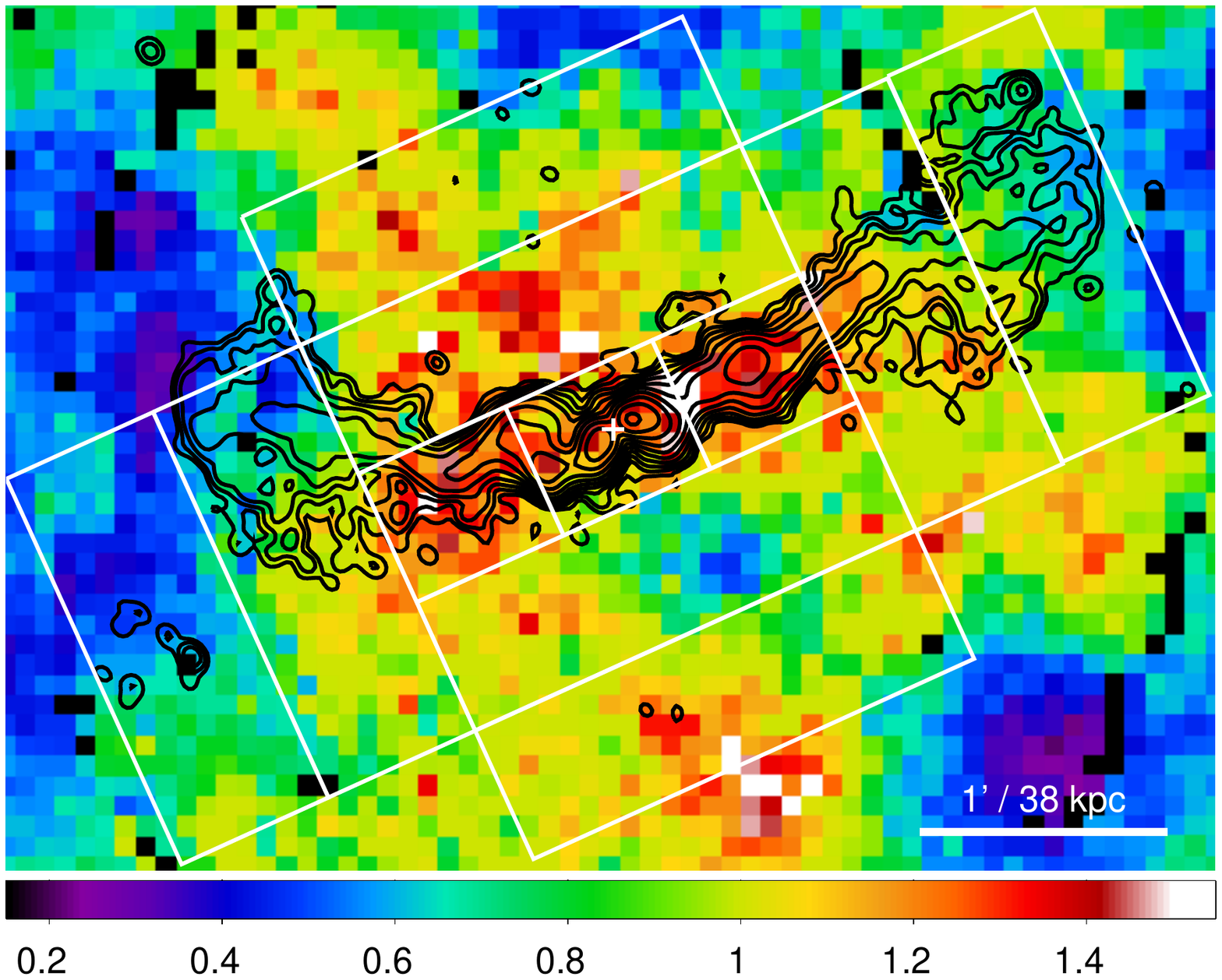}}
\caption{\label{fig:mapzoom} Abundance map of the core of AWM~4, with GMRT
  610~MHz contours overlaid. Rectangular regions were used to examine the
  variation in abundance across and along the jet. The white cross marks the position of the radio core.}
\end{figure}

\begin{figure*}
\centerline{\includegraphics[width=\textwidth,viewport=36 274 577 518]{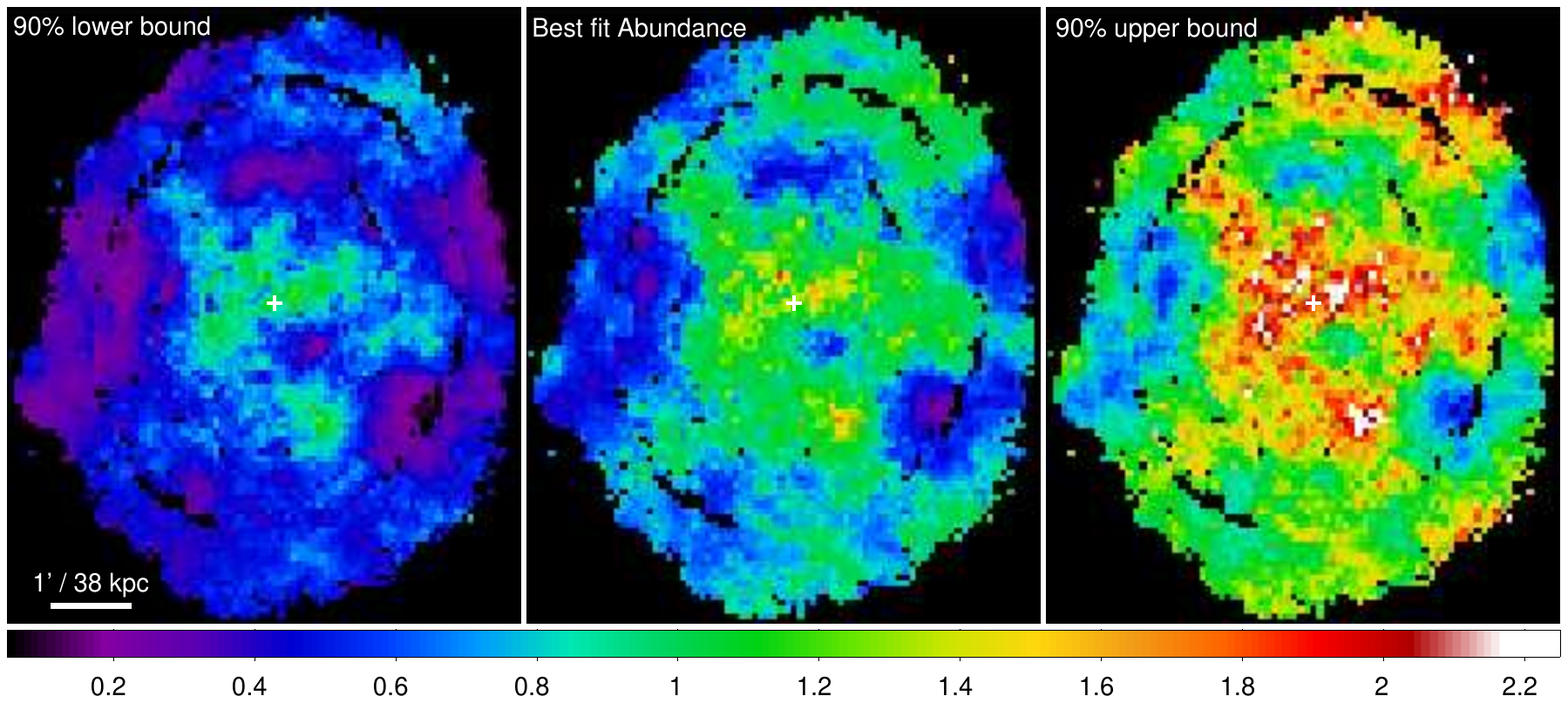}}
\caption{\label{fig:Zerr} Abundance maps of AWM~4 in solar units, showing
  (\textit{centre}) the best-fitting value and the 90\% lower
  (\textit{left}) and upper (\textit{right}) bounds. Abundances are in
  solar units. Note that the best-fit abundance map is the same as that
  shown in Figure~\ref{fig:maps}, but that the colour scale has been
  selected to allow direct comparison with the uncertainty maps. White crosses mark the position of the radio core.}
\end{figure*}

\begin{figure}
\centerline{\includegraphics[width=\columnwidth,viewport=20 200 560 740]{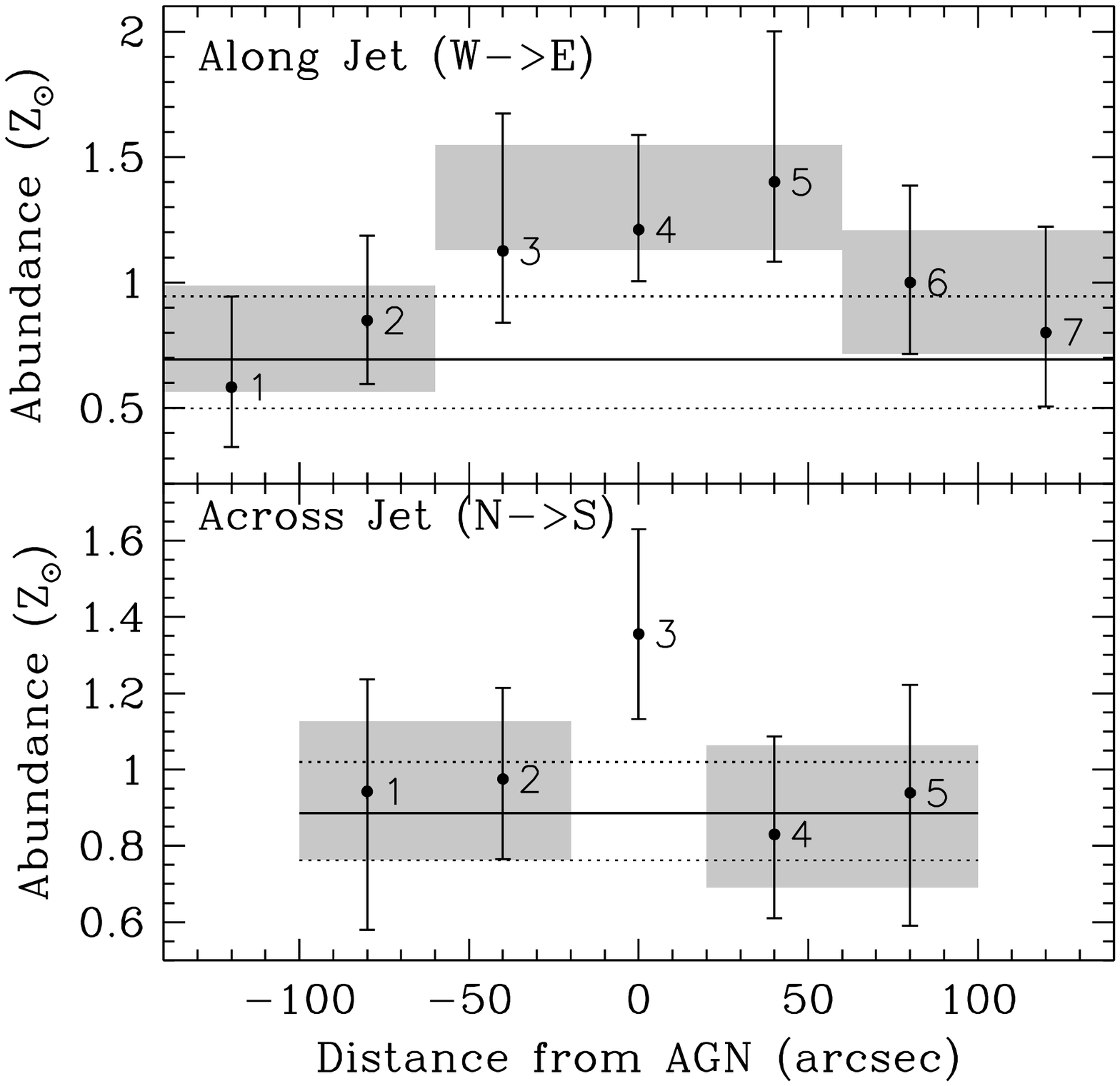}}
\caption{\label{fig:jetZ} Measured abundances in profiles running east to
  west along the radio jets and north to south across the jets, using the
  regions shown in Figure~\ref{fig:mapzoom}. Black points with error bars
  indicate measurements in individual bins with 90\% uncertainties.
  Numerals beside each point indicate the bin numbers referred to in the
  text. Grey boxes show the uncertainty region when adjacent bins are
  fitted simultaneously. Black solid lines represent the combined best fit
  to spectra 1 and 7 on the E--W profile, and 1, 2, 4 and 5 on the N--S
  profile, with 90\% uncertainties shown by dotted lines. }
\end{figure}

To examine the high abundances associated with the radio jets, we placed a
number of rectangular regions along and across the jet, shown in
Figure~\ref{fig:mapzoom}. Smaller regions are used in the inner part of the
jet to allow us to look for any central abundance peak, larger regions
outside to minimise the uncertainties on abundance. Spectra were extracted
from these regions and fitted with an absorbed APEC model. The resulting
abundances are shown in Figure~\ref{fig:jetZ}. The east to west profile uses
the two large rectangular regions at each end, and the smaller rectangles
along the jet; the north to south profile compares the upper and lower pairs of
large rectangular regions, and the central region comprising the three
small rectangles combined. 

While the abundances in neighbouring regions are comparable, there is a
clear trend for higher abundances in the inner jets (the three central
regions of the east--west profile) and declining abundance outside that
area. The abundance of the westernmost region is lower than the abundances
in the inner jet at 90\% significance. Combining regions of similar
metallicity, we find that the inner part of the jets (regions 3-5 of the
E--W profile, or region 3 of the N--S) is more enriched than the regions at
the eastern end of the jet at 3.2$\sigma$ significance, but only at a
2.0$\sigma$ level in comparison to the western regions.  However, comparing
the inner jet to a combination of the extreme western and eastern regions
shows a 3.4$\sigma$ difference.  The northern and southern regions,
combined in pairs, are less abundant at the 2.4-2.7$\sigma$ level, or
3$\sigma$ if all four are simultaneously fitted.  In general, we conclude
that the high abundance region is more extended E--W than N--S, following
the jet, and that its abundance is significantly greater than its
surroundings, by $\sim$0.4\Zsol.

\subsection{Accuracy of the spectral maps and potential sources of bias}
\label{sec:bias}

\begin{figure}
\centerline{\includegraphics[width=\columnwidth,viewport=30 450 565 740]{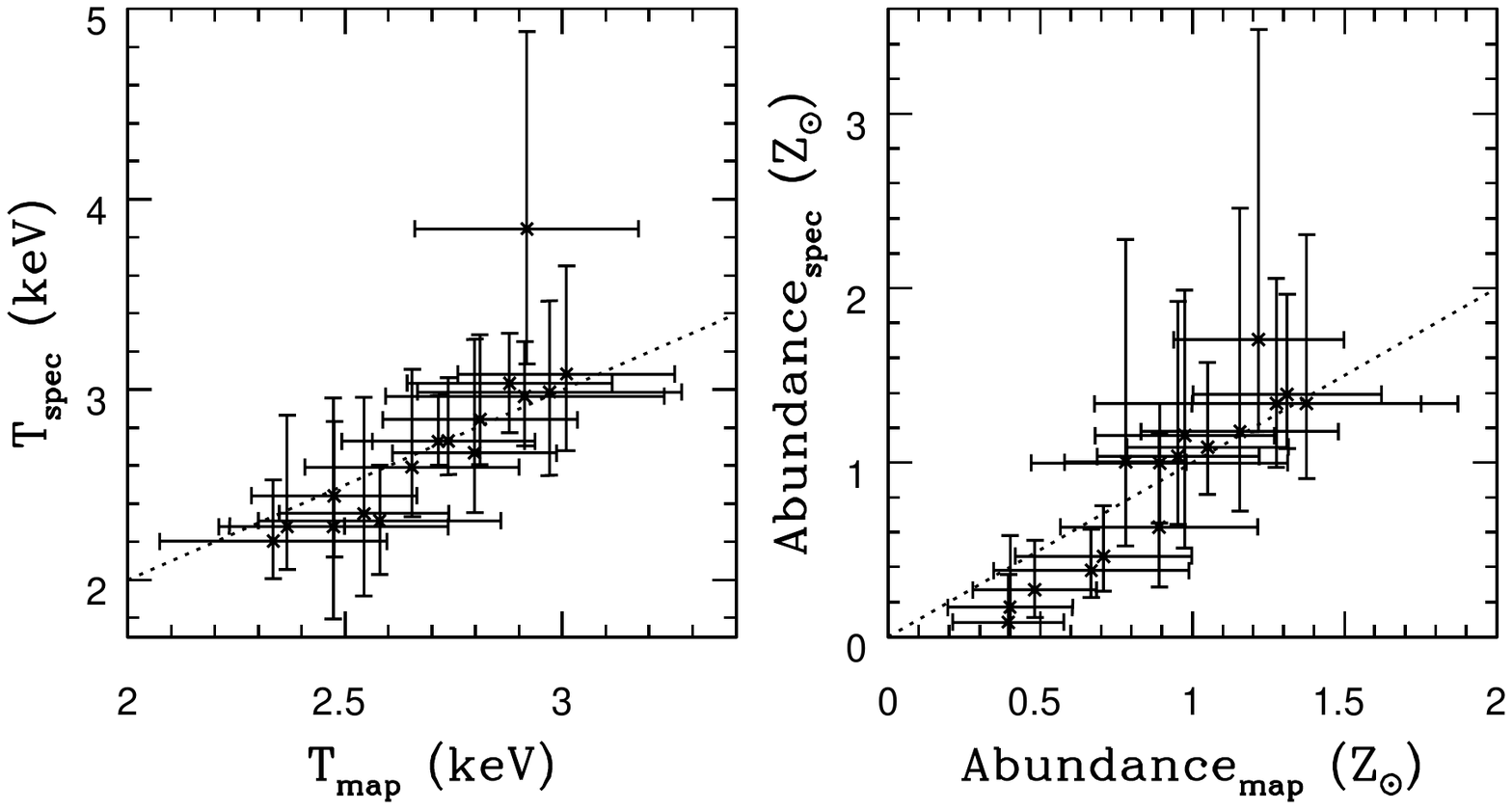}}
\caption{\label{fig:maptest}Comparison of temperatures and abundances
  derived from the spectral maps and from normal spectral fitting for the
  numbered regions shown in Figure~\ref{fig:maps}. Error bars indicate the
  90\% uncertainty on the parameter from the spectral fits, or the range of
  parameter values found in the region in the ``best-fit'' spectral maps.
  Dotted lines indicate agreement between the two methods.}
\end{figure}

To test the accuracy of the maps we defined regions covering specific
temperature and abundance features, extracted spectra from these regions,
and fitted them. The regions contain between $\sim$660 and $\sim$2900 net
counts in the 0.7-7.0~keV band.  While the spectral extraction and fitting
process is identical in mapping and normal spectral analysis, these regions
were not constrained to contain a fixed number of counts, so should provide
a test of the smoothing-like effect of the mapping process. It also allows
us to determine how well the variation within map regions corresponds to
the uncertainty on the normal spectral fit.  Figure~\ref{fig:maptest} shows
comparisons of the range of temperatures and abundances found in the map
regions with the values derived from the spectral fits.

In general the maps appear to provide an accurate estimate of both
temperature and abundance. One spectral fit, for region A, finds a
significantly higher temperature than the map suggests. This appears to be
a smoothing issue, the spectral extraction regions used to create the map
being significantly larger than region A, which contains only
$\sim$900 net counts (0.7-7.0~keV). The abundance measurements typically
have larger uncertainties, and a greater variation in map pixel values,
particularly in the highest abundance regions. The highest abundance of any
spectral fit is found in region B. The region contains $\sim$1200
net counts, again suggesting that smoothing lowers the maps values, but the
spectral fit agrees with the map within the uncertainties. The regions with
lowest abundances (regions C,D,E,F and G) appear to have map
values which are slightly overestimated. However, they are again consistent
within the uncertainties.

The abundances measured in the map spectral fits could be biased by a
number of issues which arise from the assumption of a simple single
temperature plasma model when fitting spectra which must, on some level, be
produced by gas whose properties vary within the spectral extraction
region.  One possible problem is the inverse Fe-bias
\citep{Rasiaetal08,Simionescuetal09,Gastaldelloetal10}, which can affect
abundance measurements made using single temperature models fitted to gas
with a mean temperature of 2-4~keV. If the gas has a complex,
multi-temperature structure, abundance may be overestimated owing either to
the mixture of differing levels of Fe-L and Fe-K shell emission along the
line of sight \citep{Rasiaetal08}, or the behaviour of the spectral fitting
code when a single temperature model is used to describe a
multi-temperature spectrum \citep{Gastaldelloetal10}.

Testing for this type of problem is difficult, in that high
signal--to--noise spectra are required if multi-temperature models are to
be fitted successfully, but larger regions are more likely to contain
significant temperature variations. To determine whether the inverse Fe-bias is
likely to affect our abundance measurements, we performed three tests:

\begin{enumerate}
\item We combined regions 3-5 of the E--W profile described in
  Section~\ref{sec:metals} and fit a two-temperature plasma model to the
  0.7-7~keV band, with the two abundances tied. The model converges on
  temperatures of $\sim$2.6~keV for both components, but neither
  temperature or normalisation are constrained, indicating that either
  component can fit the data adequately without need for a significant
  contribution from the other. The abundance is found to be
  1.33$^{+0.22}_{-0.20}$\Zsol, in good agreement with the single
  temperature fit. The gas thus appears to be close to single-temperature
  in this region.
\item Using the same combined spectra from regions 3-5, we fitted
  single--temperature models separately to the 0.7-2~keV and 2-7~keV bands.
  Table~\ref{tab:bands} shows the results of these fits. The hard band fit
  finds a somewhat lower abundance, while the soft band fit finds higher
  abundance and temperature. However, the uncertainties are large and both
  abundances are consistent with the broad--band best--fitting value and
  with each other at the 90 per cent level.
\item We recreated the spectral maps using $\sim$10\arcs\ pixels, requiring
  3200 counts per spectrum, and allowing spectral extraction regions of
  radius 19-74\arcs. The spectra were fitted using a two-temperature APEC
  model with abundance tied between the two components. These fits were
  successful for $\sim$50 per cent. of the spectra, with a single component
  dominating in the remainder. The regions in which the two--temperature
  model produces a useful fit are not correlated with the X--ray or radio
  surface brightness, or with the temperature or abundance structures. The
  resulting abundance map shows the same structure as our main abundance
  map, and the differences in abundance between the two maps are not
  correlated with any particular regions, i.e.  we see no evidence of a
  systematic over-- or under--estimation of abundance along the radio jets.
  There is no indication that those pixels with a significant contribution
  to the spectrum by the second temperature component have higher or lower
  abundances in the two-temperature fits. The only region near the radio
  source in which we see both a high abundance and evidence that a
  two-temperature model is a reasonable fit to the data is in the knot of
  the western radio jet (roughly corresponding to region H on
  Figure~\ref{fig:maps}).  Spectra from pixels in the two highest abundance
  regions outside the jets (regions I and K) appear to be best modelled as
  single-component spectra, as do some of the lowest abundance pixels
  (region C). The low abundance clump just south of the jet, in the east of
  region J, includes a mix of pixels whose spectra produce useful
  two--temperature fits and those which are best fit with single
  temperature models. In general we conclude that there is no evidence that
  the abundance distribution would be significantly altered by the use of
  two--temperature fits.
\end{enumerate}

\begin{table}
\begin{center}
\begin{tabular}{lcccc}
Energy band & $kT$  & Abundance & red. $\chi^2$ & d.o.f. \\
            & (keV) & (\Zsol)   &               &        \\
\hline\\[-3mm]
0.7-7.0 & 2.66$^{+0.10}_{-0.09}$ & 1.32$^{+0.23}_{-0.19}$ & 0.926 & 327 \\
0.7-2.0 & 2.92$^{+0.34}_{-0.29}$ & 1.63$^{+0.54}_{-0.37}$ & 0.941 & 215 \\
2.0-7.0 & 2.72$^{+0.24}_{-0.22}$ & 1.08$^{+0.51}_{-0.43}$ & 0.848 & 104 \\
\end{tabular}
\end{center}
\caption{\label{tab:bands} Spectral fits to the broad band emission from
  the region along the radio jets, compared with fits to soft and hard sub--bands.}
\end{table}

A powerlaw emission component associated with the radio jets or lobes could
also affect the measured abundance. If the powerlaw slope is similar to
that of the bremsstrahlung continuum, fitting the spectrum with a simple
plasma model is likely to result in an underestimate of abundance. If the
powerlaw provides more emission at higher energies, the plasma model
temperature is likely to be biased high, and abundances overestimated. The
expected flux from inverse Compton scattering in the radio lobes is too low
to be detected, and indeed no evidence powerlaw emission is found (see
paper I). To test whether there could be significant powerlaw emission from
the jets, we fit apec+powerlaw models to the spectra extracted in the E--W
and N--S profiles. If a powerlaw component were associated with the jets we
would expect to see the highest fluxes from regions 3-5 of the E--W
profile, and none in regions 1, 4 and 5 of the N--S profile. In practise we
find powerlaw contributions to be consistent with zero in all regions
except regions 2 and 3 of the E--W profile, and region 1 of the N--S
profile. In these regions the best-fitting powerlaw index is inverted, and
the component only produces significant additional flux above 5 keV. If the
powerlaw index is fixed at $\Gamma$=2, as expected from the radio spectral
index (see paper I), the powerlaw flux is consistent with zero. We
therefore conclude that no significant powerlaw emission is detected from
the jet, and that the abundances are not biased in this way.

High abundances could also be found in error if the plasma is not in
ionisation equilibrium. \citet{Kaastraetal09} point out that the Maxwellian
electron distribution assumed for thermal plasmas is not valid in shocked
regions. Non-thermal electrons alter the X--ray spectrum and can lead to
the overestimation of the abundance; A possible example is seen in the high
abundance arc associated with the shock in HCG~62 \citep{Gittietal10}.
Both shocks or mixing of non-thermal electrons into the thermal plasma
could occur along the edge of the radio jets in AWM~4. The weak surface
brightness features associated with the jets could indicate the presence of
compressed gas (see paper I, fig.~2). However, we might expect that
shocks capable of affecting the abundance measurements would be detectable
as either density or temperature features. The only regions where such
features may be visible are in the western knot and eastern bend of the
jets, but our data are insufficiently deep to confirm or reject this. 

We also note that we might expect to find both mixing of relativistic
electrons from the radio source and multi-phase gas in the lobes of the
radio source as well as the jets, since the lobes appear to contain a mix
of thermal and relativistic plasmas (see paper I). The lack of
high abundances coincident with the lobes therefore argues against these
sources of bias being effective. The small extension of the high abundance
region north of the galaxy core also cannot be explained as a shock or
jet-related feature.

We therefore conclude that while it is likely that the spectra used in the
maps do contain emission from multiple plasma components with different
temperatures and abundances, the variation is probably not large enough to
affect our results, and the single-temperature fits provide a sufficiently
accurate estimate of abundance for our purposes.

\subsection{Comparison with XMM-Newton spectral maps}
Comparison of the \chandra\ spectral maps with the \xmm\ maps of OS05
reveals some differences, particularly in the abundance distribution.  The
\xmm\ abundance map contains a ridge of high metallicity extending from the
galaxy core northwest along the northern edge of the western jet. Testing
with normal spectral fits to the \xmm\ data confirmed that this region had
enhanced abundances, ruling out a problem specific to the mapping
technique. OS05 suggested that the feature might arise from stripped
material or a shock associated with the jet.  The \chandra\ data show no
such ridge, though high abundances are seen in the galaxy core.

To confirm the lack of such a feature in the \chandra\ data we extracted a
spectrum in an annular segment defined from the \xmms\ abundance map,
extending northwest, from $\sim$45-120\arcs\ from the AGN, equivalent to
region 1 of Figure~9 and Table~5 of OS05. This spectrum, and the \xmms\
spectra extracted from region 1 by OS05 using SAS 6.0, were then fitted
with an absorbed APEC model using a range of energy bands.

We find that 1) although \chandra\ ACIS and \xmms\ EPIC-pn and MOS cameras
give consistent abundances within the errors, the EPIC-pn spectra produce
the highest best-fit abundances and ACIS the lowest, and 2) the abundance
measured is affected by the energy band used, with wider energy bands
producing higher abundances in this region. Using our 0.7-7.0 keV energy
band, EPIC-MOS produces best-fit abundances very similar to ACIS
(0.78$^{+0.39}_{-0.28}$ compared to 0.70$^{+0.17}_{-0.15}$ \Zsol) while the
EPIC-pn abundance is significantly greater (1.01$^{+0.28}_{-0.24}$. This
suggests that \xmms\ map feature is either the product of inaccurate
calibration of the EPIC-pn, or that it is caused by the presence of some
spectrally hard emission component which affects the EPIC-pn more than ACIS
or EPIC-MOS because of its greater effective area at high energies.

As discussed in OS05, the high abundance ridge does not correspond to
detector structure on either EPIC-MOS or -pn. The ridge is located on the
focal--point CCD of EPIC-pn, at least 45\arcs\ northwest of the CCD edge at
the centre of the array. There are no structures in the background data
files at the position of the ridge.  It therefore seems unlikely that a
calibration problem associated with a particular CCD feature could be the
cause of the high abundances.

Examining hard band images (5-7 keV) some clumps of emission are visible,
but these are not consistent between instruments and involve only a handful
of counts. Unfortunately neither dataset is sufficiently deep to determine
whether an excess is present. Fitting the MOS and pn spectra described
above using an apec+powerlaw model, we find neither can constrain the
powerlaw index, and that for various fixed index values the MOS abundance
is still consistently lower than that measured for the EPIC-pn. This argues
against an additional spectral component as the cause of the ridge, and
leaves the origin of the feature uncertain.  To check whether any hard
emission affects our \chandra\ maps we recreated the maps using 3200 count
spectra, and ignoring energies above 3.0~keV. We find that the map
structures do not change, suggesting that our results are not affected by
extraneous hard emission.

\section{Discussion}
\label{sec:discuss}

While we have focussed on the interaction between the AGN and ICM, there
are other mechanisms which could affect the abundance and temperature
structure of AWM~4. Gas motions associated with mergers and interactions
are one possibility. The overall structure of the temperature map, with
higher temperatures to the south of the radio jet than to its north at
moderate radii, with the pattern reversed at large radii, could in
principle be an indication of gas sloshing. The infall of a subcluster can
disturb the gravitational potential of the cluster, causing the cluster
core to oscillate around its centre, sloshing the gas of the core back and
forth.  Examples of gas sloshing have been documented in a number of
clusters
\citep[e.g.,][]{Markevitchetal01,MazzottaGiacintucci08,Simionescuetal10},
and confirm the results of numerical simulations
\citep[e.g.,][]{AscasibarMarkevitch06}, in that the sloshing is observed to
produce a spiral pattern of cold fronts, where cool gas from the cluster
core is moved outward and brought into contact with hotter material. The
staggered pattern of temperature reversals seen to north and south of the
radio source could be produced by such spiral sloshing.  The apparent
southward motion of NGC~6051 could also be explained if the cluster core is
in motion relative to its surroundings. Sloshing provides a mechanism for
transporting metals out of the cores of clusters \citep{Simionescuetal10},
and would therefore be relevant to our study of AWM~4.

However, there are several strong arguments against sloshing being
  important in this cluster: 
\begin{itemize}
\item No cold fronts or sharp changes in surface brightness are detected
  anywhere in AWM~4, and there is no evidence of surface brightness
  features corresponding to the temperature structure.  There is evidence
  that the surface brightness is generally greater north of the radio
  source than to the south, but this appears to be caused by a mild offset
  between the position of NGC~6051 and the centroid of the ICM emission. In
  general the surface brightness suggests that the ICM is relaxed.
\item There is no obvious candidate for the subcluster needed to
  set the sloshing in motion. No substructure is observed in the X--ray, no
  galaxies of comparable luminosity to NGC~6051 are observed in the system,
  and there is no evidence of substructure in the galaxy spatial or
  velocity distribution \citep{KoranyiGeller02}. 
\item If sloshing has occurred, we would expect correlation between
  temperature and abundance structure. Instead, we find NGC~6051 located
  near the centre of the inner, near--solar abundance region, with high
  abundances correlated with low temperatures north of the radio jet, but
  with high temperatures south of the jet.
\end{itemize}
We therefore conclude that while NGC~6051 may be in motion, it is unlikely
that large--scale gas sloshing is occurring in AWM~4.

\subsection{Metal mass and energetics}
Under the assumption that iron is the dominant element determining the
measured metallicities, we can estimate the mass of iron required to
produce the raised abundance seen along the jets,

\begin{equation}
m_{Fe}=\rho V\Delta Z\gamma_{Fe},
\end{equation}

where $\rho$ is the density of the gas in the jets, V is the volume of the
region, $\Delta Z$ is the excess abundance above the background enrichment
level and $\gamma_{Fe}$ is the solar iron mass fraction
\citep[1.28$\times10^{-3}$ for the abundances of][]{GrevesseSauval98}.  For
this purpose, we define the volume of high enrichment along the jets using
the smallest rectangular regions in Fig.~\ref{fig:mapzoom} and assuming a
rotational symmetry around the jet axis, to produce three cylinders of
radius 11~kpc and length 20~kpc. Gas densities in each cylinder are taken
to be the average density at the mean radius of the region, as determined
in paper I. We take the abundance of this region to be 1.3\Zsol, based on
the mean value found for the regions along the jet. The abundances measured
from the rectangular regions around the jet suggest a background level of
0.9\Zsol, but the clumpiness of the abundance distribution on these scales
makes this uncertain. To get a more reliable measure of the general level
of enrichment in the central part of the cluster, we extracted a spectrum
from a circular region of radius $\sim$100\arcs\ ($\sim$65~kpc) centred
just south of the radio core, with the region of the jets excluded. This
encloses the whole high abundance part of the cluster core, and while there
is spatial variation in metallicity, it should provide a reasonable average
enrichment level.  The best fitting abundance for this region is
0.86$^{+0.09}_{-0.08}$\Zsol, suggesting that a background abundance of
0.9\Zsol is acceptable. This then suggests an excess iron mass of
$\sim1.4\times10^6$\Msol.

Beyond the 0.9\Zsol central region the mean abundance falls to
$\sim$0.6\Zsol. From the gas mass profile of OS05, we estimate the total
gas mass within 65~kpc to be 6.3$\times10^{10}$\Msol. For an excess
abundance of 0.3\Zsol, this suggests that the volume has been enriched by
an additional 2.4$\times10^{7}$\Msol\ of iron.

NGC~6051 is the most massive galaxy in AWM~4 by a significant margin, and
given its central position it is likely to be the source of a large
fraction of the enrichment. If the super-solar abundances are associated
with material entrained by the radio jets, the gas now seen along the jets
must have been enriched in the galaxy core. Using the regions described
above, we can estimate that if the enriched material was originally all in
the central $\sim$10~kpc of the galaxy, $\sim$45\% of the metals originally
formed in this central region have been transported out along the jets.

We can assume that the emission observed from the region of the jets
represents a mix of gas, both highly enriched uplifted material and the
lower metallicity gas which occupied the whole region. The two phases may
be physically mixed or separate, but the measured abundance represents an
average of the two. The mass of enriched gas which must be uplifted and the
energy required to do so will depend on the level of enrichment and on the
abundance of the gas it mixes with. If the core were very highly enriched,
even a small amount of gas transported outward would produce the abundances
we observe, when mixed with the less enriched gas at that radius. We can
define the mass of gas which must be transported to be:

\begin{equation}
m_{trans.}=\frac{Z_{obs}-Z_{prior}}{Z_{enrich}-Z_{prior}} \rho V,
\end{equation}

where $\rho V=m_{gas}$=1.19$\times10^9$\Msol\ is the mass of gas in the
jets outside the core, $Z_{obs}$ is the abundance now observed in the jets,
$Z_{prior}$ is the abundance in the jets prior to mixing with more enriched
material from the core, and $Z_{enrich}$ is the abundance of that highly
enriched material. If we assume that the jets expanded into material with
abundance $Z_{prior}$=0.9\Zsol, as for the surrounding gas, we can estimate
$Z_{enrich}$ by assuming that the excess metals now observed in the jets
were once in the core. This suggests that the uplifted enriched material
had an abundance $Z_{enrich}$=1.6\Zsol. In this case
$m_{trans.}$=0.57$\times m_{gas}$.  The energy required to uplift the gas
from the core is simply the change in gravitational potential energy as the
gas is lifted from the core to the mean radius $R$,
$E=GM_{tot}(<R)m_{trans.}/R$ where $R\sim20$~kpc,
$M_{tot}(<R)$=1.54$\times10^{12}$\Msol\ is the total mass within this
radius (from OS05), and $G$ is the gravitational constant. For
$m_{trans.}$=0.57$\times m_{gas}$, this gives $E=4.5\times10^{57}$~erg.
The maximum energy which could be required ($E=7.9\times10^{57}$~erg)
corresponds to the case where the gas uplifted has $Z_{enrich}=Z_{obs}$, in
which case all the gas currently in the jets has been uplifted, displacing
that which was there prior to the outburst. Less energy will be required if
higher enrichment levels are assumed in the core; taking the 90 per cent.
upper bound on the current abundance there (1.6\Zsol) and again assuming
the excess metals in the jets originated in the core, its abundance would
have been $\sim$1.9\Zsol, giving $m_{trans.}=$0.4$\times m_{gas}$ and
uplift energy $E=3.2\times10^{57}$~erg. In any case, the energy required to
uplift the enriched material is a significant fraction of the expected
mechanical power output of the jets, which we estimate to be
9.4$\times10^{58}\phi$~erg, where $\phi$ is the filling factor of the lobes
(see paper I). Our best estimate of filling factor, for the western lobe
where a cavity is detected, is $\phi$=0.21, and the 3$\sigma$ upper limit
on the mean filling factor of the two lobes is $\phi<$0.6. This suggests
that $>$20 per cent. of the mechanical energy of the jets could be required
to produce the observed uplift.

Further limits on the level of enrichment of uplifted material can be
estimated based on surface brightness. If we assume that the pressure of
the uplifted material decreases as the gas rises, maintaining pressure
equilibrium with its surroundings, we can consider two extreme cases;
either the pressure decrease is achieved through density decrease
(expansion) or through temperature decrease. For each case, we can estimate
the expected surface brightness increase for a given abundance and mass of
uplifted gas ($Z_{enrich}$ and $m_{trans}$).

As discussed in paper I, there is some apparent increase in surface
brightness along the jets, but these features are not statistically
significant. Determining a `background' level of surface brightness is also
difficult, since the mild offset between the position of NGC~6051 and the
X--ray centroid of the ICM means that surface brightness north of the jets
is systematically brighter than to the south, and there is some evidence of
additional structures north of the jet. We determine the 0.7-3~keV surface
brightness in a number of segments of an elliptical annulus chosen to match
the mean ellipticity of the diffuse emission, corrected using a 1.05~keV
monoenergetic exposure map. As expected, surface brightness declines from
north to south across each jet by a factor larger than the statistical
uncertainty, and there is significant bin--to--bin variation around the
annulus, beyond that expected for a smooth distribution.  For excess
surface brightness to be detected in the jets, it would have to exceed that
of the brighter neighbouring bin by a significant margin. We define this as
an increase of three times the mean variation between bins across the jets.

On this basis, we can consider two possible scenarios, depending on whether
the uplifted material expands adiabatically or isothermally as it rises.
The first case would occur if conduction between the material and its
surroundings is strongly suppressed, the latter if conduction were highly
efficient. For adiabatic expansion, we can determine the expected change in
temperature for a monatomic ideal gas using the relation
T$_2$=T$_1$(P$_2$/P$_1$)$^{(\gamma-1)/\gamma}$, where T and P are temperate
and pressure before (1) and after (2) expansion, and the adiabatic index
$\gamma$=5/3. The known change in pressure then allows us to calculate the
expected change in density. We find that the uplifted material would still
be both cooler and denser than the surrounding gas, and therefore
significantly more luminous than the surrounding ICM with even 1.3\Zsol\
abundance. The surface brightness limit for the west jet rules out this
scenario. For the east jet, the upper limit on surface brightness suggests
limits on the uplifted material of $Z_{enrich}\la$1.4\Zsol and
$m_{trans}\la$0.8$\times m_{gas}$. For the second case, in which the
uplifted gas expands isothermally as it rises, we estimate limits of
$Z_{enrich}\la$2\Zsol\ and $m_{trans}\la$0.36 in the western jet, and
$Z_{enrich}\la$1.7\Zsol and $m_{trans}\la$0.5 in the eastern jet. In
practice, this suggests that the gas mainly expands as it rises, with only
slight cooling. These limits ignore the possible presence of channels or
sub-cavities from which ICM plasma has been excluded by the jets, which
would reduce surface brightness.  Nonetheless, we conclude that our initial
estimate of the mass of uplifted gas and the energy required to transport
it are probably correct.

We can estimate the timescale over which these levels of enrichment might
occur using the relation of \citet{Bohringeretal04},

\begin{equation}
t_{enr} = (10^{-12}S\nu_{\rm Fe} + 1.5\times10^{-11}t_{15}^{-1.3}Z_*\gamma_{
  Fe})^{-1}\frac{M_{\rm Fe}}{L_B/L_{B\odot}},
\end{equation}

where $t_{enr}$ is the timescale required to produce the iron, $S$ is the
supernova rate \citep[we assume 0.18 SNu or $\sim$0.0125 SN yr$^{-1}$ for
NGC~6051, from][]{Cappellaroetal99}, $Z_*$ is the stellar metallicity,
$\nu_{\rm Fe}$=0.7\Msol\ is the iron yield from SNIa \citep[][ for the WDD2
model]{Nomotoetal97} and $\gamma_{Fe}$ is the iron mass fraction in stellar
mass loss. The rate of stellar mass loss is derived from the approximation
$\dot{M}=1.5\times10^{-11}t_{15}^{-1.3}L_B/L_{B\odot}$ for a passively
evolving old stellar population \citep{Ciottietal91}, where $t_{15}$ is the
age of the stellar population as a fraction of 15~Gyr. Neither the age or
mean abundance of the stellar population of NGC~6051 have been measured,
and we therefore assume an age of 10~Gyr (equivalent to all stars forming
at a redshift $\sim$2) and an abundance of 1.6\Zsol\ \cite[comparable to
similar ellipticals,][]{Denicoloetal05}. On this basis, enrichment of the
$\sim$0.9\Zsol\ region from a starting abundance of 0.6\Zsol\ would require
$\sim$1.9~Gyr. Unsurprisingly, this is far longer than the estimated
timescale of the current AGN outburst \citep[$\sim$170~Myr,][paper
I]{Giacintuccietal08}. We would expect metals in the inner part of the
cluster to have built up over the whole lifespan of the system, and while
we are treating this as a plateau level of enrichment, it is consistent
with the broad abundance peak seen in most galaxy clusters (see paper I for
a radial abundance profile).  However, this estimate suggests that it is at
least plausible that NGC~6051 has provided most or all of the enrichment in
this region.

The mass of iron required to enrich the gas along the jet from 0.9 to
1.3\Zsol\ could be produced in 107~Myr, well within the outburst timescale.
Since the enrichment of the $\sim1.3$\Zsol\ gas is occurring in a reduced
volume at the core of the galaxy, this timescale is probably
underestimated, but considering that the stellar density is highest in this
region it seems reasonable to expect that enrichment by stellar processes
is capable of producing the high abundances observed along the jet.

The location of the enriched gas along the jets rather than in the lobes
provides some information about the method by which the gas is transported.
In the nearby radio galaxy Centaurus A, it has been suggested that material
can be directly entrained from stars located within radio jets
\citep{Nulsenetal10}. While this provides a natural route for increasing
the mass loading of jets to the levels required to provide pressure
equilibrium between radio lobes and their environment (see paper I for
discussion of this possibility in AWM~4), it is unlikely
to provide the abundance structures we observe. We would expect material
entrained within the jets themselves to be rapidly transported into the
lobes, and probably to have a high temperature and low density,
precluding detectable X--ray emission in the energy band which we have
observed. 

The more likely alternative is that the enriched gas is being uplifted more
slowly by gas motions associated with the growth of the radio jets and
lobes. Numerical modelling of AGN jet/gas interactions suggest that the
buoyant rise of cavities associated with radio lobes causes subsonic gas
motions, drawing gas out of the cluster core along the line of the jets
\citep[e.g.,][]{Bruggen02, Roedigeretal07}. In this scenario, the enriched
material occupies a volume around the jets, where gas motions produce an
overall outward motion similar to the growth velocity of the radio source,
but without being accelerated by interaction with the relativistic jet
plasma. The lack of shocks associated with the lobes of 4C+24.26 indicates
that they are moving subsonically, and the buoyant timescale
($\sim$130~Myr, paper I) is comparable to the radiative age of the lobes.
It is therefore likely that the lobes have risen buoyantly to their current
position, uplifting enriched gas behind them.

\subsection{Other sources of enrichment}

In principle, metals could be produced along the jets through local star
formation, perhaps caused by compression of cool material by the jets.
Assuming that our mean metallicity measurements are determined largely by
the iron abundance, we can estimate that $\sim$2$\times10^{7}$
SNII would be needed to enrich the gas to 1.3\Zsol, assuming an initial
0.9\Zsol\ abundance.  With a simple assumption of one supernova per
100\Msol\ of stars formed, this is equivalent to a star formation rate
(SFR) of $\sim12~{\rm M}_\odot~{\rm yr}^{-1}$ over the 170~Myr timescale
of the AGN outburst.

Assuming a 10\% efficiency rate for the star formation, we require
$2\times10^{10}$\Msol\ of cool gas to have been present along the jets, and
presumably larger masses in the galaxy as a whole. This is comparable to
the mass of the hot ICM within a radius equal to the extent of the radio
source ($\sim6\times10^{10}$\Msol, OS05).  Cool gas and star formation is
detected in some clusters
\citep[e.g.,][]{EdgeFrayer03,SalomeCombes03,Bildfelletal08,Pipinoetal09},
but is typically associated with large cool cores or mergers. The estimated
SFR is comparable to those observed in some massive cooling flow clusters
\citep{O'Deaetal08} and large compared to rates found from UV observations
of less extreme systems \citep{Hicksetal10}. If the enrichment is produced
by SNII, we would expect to see a clear change in abundance ratios between
regions inside and outside the jet, and probably considerably higher
abundances of Si than Fe. We fitted vapec models to the spectra
from the regions shown in Figure~\ref{fig:mapzoom}, leaving Fe and Si free
to vary.  The resulting abundances are in many cases not very tightly
constrained, but we generally find abundance ratios Si/Fe$\sim$1, with
uncertainties of a factor of 2. These uncertainties are large enough to
render estimates of the numbers of supernovae of different types necessary
to enrich the gas impractical, but there is no systematic change in the
ratio outside the jets, suggesting that the gas throughout the core has
probably been enriched by a similar mix of SNIa and SNII.  The alignment of
the enriched region along the jet, and its extent well beyond both the
central corona and the optical body of the galaxy, is also difficult to
explain through star formation fuelled by cooling or gas brought into
NGC~6051 by a merger.  We might also expect the $\sim2\times10^9$\Msol\ of
young stars formed within the lifetime of the jet to be detectable, as a
significant fraction would form at large radii along the galaxy minor axis.
It therefore appears unlikely that the high abundances observed in the jet
are the product of recent star formation.

\section{Summary}
\label{sec:sum}
Temperature and abundance maps of the poor cluster AWM~4 reveal a high
degree of structure in this relatively relaxed poor cluster.  Features in
both temperature and abundance are found to correlate with the jets of the
central radio source, with a cool region corresponding to the eastern lobe
and cavity, and supersolar abundances extending from the galaxy core along
both jets. Testing against normal spectral analysis, and variation of the
number of counts and energy band used in the map spectral fits shows the
maps to be reliable, and we conclude that the features correspond to real
physical structures within the ICM.

The location of high abundances along the jets suggests that material
enriched in the inner parts of NGC~6051 has been entrained and is being
transported out of the galaxy, roughly along its minor axis. The mass of
iron required to produce such a feature ($\sim1.4\times10^6$\Msol, assuming
enrichment by 0.4\Zsol) is relatively modest, and it is likely that it
could be produced in the central region of NGC~6051 on a timescale
comparable to that estimated for the AGN outburst. The energy required to
transport the gas to its observed location is $\sim3-8\times10^{57}$~erg,
depending on the abundance of the uplifted material and the gas it mixes
with. This is a significant fraction of the estimated total energy required
to inflate the lobes of the radio source.  An extended region of near-solar
abundances which extends out to $\sim$65~kpc in the ICM is also likely a
product of enrichment by the central galaxy, though over timescales much
longer than the AGN outburst.  Galaxy motions and previous AGN outbursts
may have contributed to transporting metals into this region.

While it is possible that some degree of bias affects the abundance
measurements, arising either from the complex temperature structure of the
ICM or from non-thermal electron populations associated with the jets,
neither possibility seems able to explain the observed abundance
structures. We therefore conclude that AWM~4 is one of the growing number
of systems in which evidence is seen for enrichment of the ICM via
entrainment of high-abundance gas by radio jets.

\medskip
\noindent{\textbf{Acknowledgements}}\\
The authors thank the anonymous referee for a number of comments which have
materially improved the paper. Support for this work was provided by the
National Aeronautics and Space Administration through Chandra Award Number
GO8-9127X-R issued by the Chandra X-ray Observatory Center, which is
operated by the Smithsonian Astrophysical Observatory for and on behalf of
NASA under contract NAS8-03060. E. O'Sullivan acknowledges the support of
the European Community under the Marie Curie Research Training Network. We
thank the staff of the GMRT for their help during the observations. GMRT is
run by the National Centre for Radio Astrophysics of the Tata Institute of
fundamental Research. We acknowledge the usage of the HyperLeda database
(http://leda.univ-lyon1.fr).

\bibliographystyle{mn2e}
\bibliography{../../paper}

\label{lastpage}
\end{document}